\documentclass[preprintnumbers, prd, showpacs, floatfix,twocolumn,
preprintnumbers, letterpaper, superscriptaddress]{revtex4}
\usepackage{amsfonts}
\usepackage{amsmath}
\usepackage{amssymb,epsf}
\usepackage{latexsym}
\usepackage{graphicx,epsfig}
\usepackage{amssymb}
\usepackage{subfigure}
\usepackage[colorlinks=true,citecolor=blue,linkcolor=blue,urlcolor=black]{hyperref}
\usepackage{epstopdf}

\graphicspath{{Images/}}

\begin{document}

\title{Reentrant phase transition of Born-Infeld-AdS black holes}
\author{Amin Dehyadegari}
\affiliation{Physics Department and Biruni Observatory, Shiraz University, Shiraz 71454,
Iran}
\author{Ahmad Sheykhi}
\email{asheykhi@shirazu.ac.ir}
\affiliation{Physics Department and Biruni Observatory, Shiraz University, Shiraz 71454,
Iran}
\affiliation{Research Institute for Astronomy and Astrophysics of Maragha (RIAAM), P.O.
Box 55134-441, Maragha, Iran}

\begin{abstract}
We investigate thermodynamic phase structure and critical behaviour of
Born-Infeld (BI) black holes in an anti-de Sitter (AdS) space, where the
charge of the system can vary and the cosmological constant (pressure) is
fixed. We find that the BI parameter crucially affects the temperature of
the black hole when the horizon radius, $r_{+}$, is small. We observe that
depending on the value of the nonlinear parameter, $\beta $, BI-AdS black
hole may be identified as RN black hole for $Q\geq Q_{m}$, and
Schwarzschild-like black hole for $Q<Q_{m}$, where $Q_{m}=1/\left( 8\pi
\beta \right) $ is the \textit{marginal} charge. We analytically calculate
the critical point ($Q_c,T_c, r_{+c}$) by solving the cubic equation and
study the critical behaviour of the system. We also explore the behavior of
Gibbs free energy for BI-AdS black hole. We find out that the phase
behaviour of BI-AdS black hole depends on the charge $Q$. For $Q>Q_{c}$, the
Gibbs free energy is single valued and the system is locally stable ($%
C_{Q}>0 $), while for $Q<Q_{c}$, it becomes multivalued and $C_{Q}<0$. In
the range of $Q_{z}<Q<Q_{c}$, a first order phase transition occurs between
small black hole (SBH) and large black hole (LBH). Interestingly enough, in
the range of $Q_{t}\leq Q\leq Q_{z}$, a reentrant phase transition occurs
between intermediate (large) black hole, SBH and LBH in Schwarzschild-type
region. This means that in addition to the first order phase transition
which separates SBH and LBH, a finite jump in Gibbs free energy leads to a
\textit{zeroth order} phase transition between SBH and intermediate black
hole (LBH) where initiates from $Q=Q_{z}$ and terminates at $Q=Q_{t}$.
\end{abstract}

\maketitle

\section{Introduction}

Phase transition is a most remarkable property in black hole thermodynamics
which can relate a gravitational system to an accessible experimental
system. The pioneering work on black holes phase transition was done by
Hawking and Page who proved that a first-order phase transition occurs
between thermal radiation and large black hole in the Schwarzschild
anti-de-Sitter (AdS) spacetime \cite{HP}. After that, Witten showed that
this phase transition can be related to a confinement-deconfinement phase
transition in the quark gluon plasma \cite{Witten}. Later, it has been
claimed that similar to\ the Van der Waals phase transition, a charged AdS
black hole undergoes a first-order phase transition between large black hole
(LBH) and small black hole (SBH) \cite{VDW1}. While similarity between
liquid-gas Van der Waals phase transition and small-large black hole phase
transition in the background of Reissner-Nordstrom-AdS (RN-AdS) spacetime
was provided by extending phase space of black hole thermodynamics \cite%
{Dolan,PV}. In an extended phase space, the variable cosmological constant
is treated as the pressure, while its conjugate quantity is the volume of
the system. In this approach the mass of black hole is treated as the
enthalpy \cite{enthalpy}. Critical behaviour of black holes in an extended
phase space have been explored in various setups \cite%
{variousBH,GBC,Hendi,Sherkat,Zou,Kamrani,Dayyani1,Dayyani2,Majhi}. Recently,
this approach has been utilized to present the emergence of superfluid-like
phase transition for a class of AdS hairy black hole in Lovelock gravity
\cite{superfluidBH}. Very recently, we disclosed a novel phase transition in
dilaton black holes, in an extended phase space where the charge of the
black hole is regarded as a fixed quantity \cite{NAAA}.

The reentrant phase transition (RPT) is composed of two (or more) phase
transition, which occurs by a monotonic variation of any thermodynamic
variable so that the initial and final phase of system are macroscopically
the same. This has previously been observed in a nicotine/water mixture \cite%
{Nwater}, granular superconductors, liquid crystals, binary gases,
ferroelectrics and gels (see \cite{RPTreview} and the references therein).
Recently, RPT has been discovered in the four-dimensional BI-AdS black hole
and higher-dimensional singly spinning Kerr-AdS \cite{BIRPT,KerrRPT}. These
black holes, in certain range of pressure, undergo a large-small-large black
hole phase transition where the latter \textquotedblleft
large\textquotedblright\ state refers to the intermediate black hole (IBH).
This RPT is accompanied by a finite jump in the Gibbs free energy which
referred to a zeroth-order phase transition. The zeroth-order phase
transition has been seen in superfluidity and superconductivity \cite{maslov}
and recently in the charged dilaton black hole \cite{NAAA}. The studies on
RPT in an extended phase space for different black holes have been carried
out in \cite{MRPT,quasiRPT,hgRPT,hairyRPT,massiveRPT,dSRPT,DRPT}.
Microscopic origin of black hole reentrant phase transitions has been
explored in \cite{MicroscopicRPT}. It was argued that the type of
interaction between small black holes near large/small transition line,
differs for usual and reentrant phase transitions. Indeed, for usual case,
the dominant interaction is repulsive whereas for reentrant case one may
encounter with an attractive interaction \cite{MicroscopicRPT}. It was also
shown that in reentrant phase transition case, the small black holes behave
like a Bosonic gas whereas in the usual phase transition case, they behave
like a quantum anyon gas \cite{MicroscopicRPT}. It is worth noting that in
all references mentioned above, the charge of black hole, $Q$, is regarded
as a fixed parameter and the cosmological constant (pressure) is treated as
thermodynamic quantity which can vary.

In a quite different viewpoint, it was argued that one can think of
variation of charge, $Q$, of a black hole and keep the cosmological constant
as a fixed parameter \cite{AAA}. According to this perspective, the critical
behavior occurs in $Q^{2}$-$\Psi$ plane, where $\Psi =1/2r_{+}$ is the
conjugate of $Q^2$ \cite{AAA}. In this approach, the thermodynamic process
is carried out in phase space where the black hole quantities are varied in
a fixed AdS background geometry. Treating the charge of black hole as a
natural external variable affects the thermodynamic behaviors of
gravitational system e.g., it can lead to interesting critical phenomena
\cite{AAA}. This alternative viewpoint for description of phase transition
and critical behavior, relevant response function clearly signifies stable
and unstable regime. In addition, it leads to small-large black hole phase
transition for charged AdS black holes, and completes the analogy with van
der Waals fluid system \cite{AAA}.

In this paper, following \cite{AAA}, we investigate thermodynamic phase
structure and critical behaviour of BI-AdS black holes in a phase space,
where the cosmological constant (pressure) is taken to be constant. It was
argued that the $(3+1)$-dimensional BI-AdS black hole implies a peculiar
phase transition in an extended phase space such as a reentrant phase
transition \cite{BIRPT}. As we shall see the BI parameter crucially affects
the temperature of the black hole when the horizon radius is small.\ We
shall analytically calculate the critical point by solving the cubic
equation which is missed in the previous studies (see e.g., \cite{Zeng}).
Also, we observe the interesting reentrant phase transition that occurs for
certain range of charge of black hole. Our work differs from previous
studies on phase transition in BI-AdS black hole \cite%
{Zeng,NORPT1,NORPT2,NORPT3,NORPT4,NORPT5,NORPT6,Dey}, in which they do not
study the reentrant phase transition of BI-AdS black holes, while in the
present work we focus on the reentrant phase transition of such system. The
investigation on phase transition of BI-AdS black holes has also been
performed at the fixed electric potential (the grand canonical ensemble) in
Ref. \cite{Fernando1}\textbf{.}

The framework of our paper is as follows. In section \ref{Review}, we review
the basic thermodynamic of BI-AdS black hole in four dimensional spacetime.
In section \ref{PT}, we study the critical behavior of BI-AdS black hole in $%
T-r_{+}$ plane with fixed cosmological constant. In section \ref{Gibbs}, we
investigate the Gibbs free energy in order to find the phase transition in
the system. We summarize our results in section \ref{conclusion}.
%%%%%%%%%%%%%%%%%%%%%%%%%%%%%%%%%%%%%%%%%%%%%%%%%%%%%%%%%%%%%%%%%%%%%%%%%%%%%%%%%%%

\section{Review on BI-AdS black hole \label{Review}}

The well known action of four-dimensional Einstein gravity in the presence
of BI nonlinear electrodynamics is \cite{Borninfeld}
\begin{equation}
\mathcal{S}=\frac{1}{16\pi }\int d^{4}x\sqrt{-g}\left[ \mathcal{R}-2\Lambda
+4\beta ^{2}\left( 1-\sqrt{1+\frac{F_{\mu \nu }F^{\mu \nu }}{2\beta ^{2}}}%
\right) \right] ,  \label{ac}
\end{equation}%
where $\mathcal{R}$ is the Ricci scalar curvature, $\Lambda $ is the
cosmological constant which relates to AdS radius as $\Lambda =-3/L^{2}$ and
the constant $\beta $ is the BI parameter with the dimension of mass that
relates to the string tension as $\beta =1/(2\pi \alpha ^{^{\prime }})$ \cite%
{ST}. Here, $F_{\mu \nu }=\partial _{\mu }A_{\nu }-\partial _{\nu }A_{\mu }$
is the electromagnetic tensor field where $A_{\mu }$ is the vector
potential. The static and spherically symmetric metric of spacetime is%
\begin{equation}
\mathrm{d}s^{2}=-f\left( r\right) \mathrm{d}t^{2}+\frac{1}{f(r)}\mathrm{d}%
r^{2}+r^{2}\mathrm{d}\Omega ^{2},
\end{equation}%
in which, $\mathrm{d}\Omega $ is the metric of an unit $2$-sphere with
volume $\omega =4\pi $ and the metric function $f(r)$ is given by \cite%
{Dey,Fernando2,Cai1}
\begin{eqnarray}
f(r) &=&1+\frac{r^{2}}{L^{2}}-\frac{m}{r}+\frac{2\beta ^{2}r^{2}}{3}\left( 1-%
\sqrt{1+\frac{16\pi ^{2}Q^{2}}{\beta ^{2}r^{4}}}\right)  \notag \\
&&+\frac{64\pi ^{2}Q^{2}}{3r^{2}}\ _{2}\mathcal{F}_{1}\left[ \frac{1}{4},%
\frac{1}{2},\frac{5}{4},-\frac{16\pi ^{2}Q^{2}}{\beta ^{2}r^{4}}\right] ,
\label{metric}
\end{eqnarray}%
where $_{2}\mathcal{F}_{1}(a,b,c,z)$ is the hypergeometric function, and $m$
is an integration constant which is related to the mass of the black hole
via $M=m/8\pi $, and $Q$ is the electric charge of the black hole per unit
volume $\omega $. The non-zero component of the gauge potential can be
written as%
\begin{equation}
A_{t}(r)=-\frac{4\pi Q}{r}\ _{2}\mathcal{F}_{1}\left[ \frac{1}{4},\frac{1}{2}%
,\frac{5}{4},-\frac{16\pi ^{2}Q^{2}}{\beta ^{2}r^{4}}\right] .  \label{gauge}
\end{equation}%
In the limit $\beta \rightarrow \infty $, the metric function (\ref{metric})
and the gauge potential (\ref{gauge}) reduce to `\textit{Reissner-Nordstrom}%
' (RN)-AdS black holes \cite{PV,AAA}. The Hawking temperature of BI-AdS
black hole is given by
\begin{equation}
T=\frac{1}{4\pi r_{+}}+\frac{3r_{+}}{4\pi L^{2}}+\frac{\beta ^{2}r_{+}}{2\pi
}\left( 1-\sqrt{1+\frac{16\pi ^{2}Q^{2}}{\beta ^{2}r_{+}^{4}}}\right) ,
\label{Tem}
\end{equation}%
where for large $\beta $, it can be expanded as%
\begin{equation}
T=\frac{1}{4\pi r_{+}}\left( 1+\frac{3r_{+}^{2}}{L^{2}}-\frac{16\pi ^{2}Q^{2}%
}{r_{+}^{2}}\right) +\frac{16\pi ^{3}Q^{4}}{r_{+}^{7}\beta ^{2}}+\mathcal{O}%
\left( \frac{1}{\beta ^{4}}\right) ,
\end{equation}%
where, $r_{+}$ is the event horizon which is defined by the largest positive
real root of $f(r_{+})=0$. Note that the first term in the right hand side
of the above expression is the RN-AdS black hole temperature and the next
term is the leading order BI correction to the temperature. The electric
potential at infinity with respect to the event horizon is given by
\begin{equation}
U=-\frac{4\pi Q}{r_{+}}\ _{2}\mathcal{F}_{1}\left[ \frac{1}{4},\frac{1}{2},%
\frac{5}{4},-\frac{16\pi ^{2}Q^{2}}{\beta ^{2}r_{+}^{4}}\right] .
\end{equation}%
\bigskip The other thermodynamic quantities associated with the Born-Infeld
black hole are \cite{BIRPT}%
\begin{eqnarray}
&&S=\frac{r_{+}^{2}}{4},\quad P=-\frac{\Lambda }{8\pi }=\frac{3}{8\pi L^{2}}%
,\quad V=\frac{r_{+}^{3}}{3}, \\
\mathfrak{B} &=&\frac{\beta r_{+}^{3}}{6\pi }\left( 1-\sqrt{1+\frac{16\pi
^{2}Q^{2}}{\beta ^{2}r_{+}^{4}}}\right) +\frac{4\pi Q^{2}}{3\beta r_{+}}
\notag \\
&&\times _{2}\mathcal{F}_{1}\left[ \frac{1}{4},\frac{1}{2},\frac{5}{4},-%
\frac{16\pi ^{2}Q^{2}}{\beta ^{2}r_{+}^{4}}\right] ,
\end{eqnarray}%
where $S$, $P$ and $V$ are entropy, pressure and volume, respectively. Also,
$\mathfrak{B}$ is a quantity conjugate to $\beta $ that is interpreted as
the BI vacuum polarization \cite{BIRPT}. It is easy to show that the
thermodynamic quantities of BI black hole satisfy the first law of
thermodynamic and Smarr formula, respectively, \cite{BIRPT}
\begin{equation}
dM=TdS+UdQ+VdP+\mathfrak{B}d\beta ,  \label{enthalpy}
\end{equation}%
\begin{equation}
M=2TS+UQ-2VP-\mathfrak{B}\beta .
\end{equation}%
It is notable that mass ($M$), charge ($Q$), entropy ($S$) and volume ($V$)
are written per unit volume $\omega $. In the next section, we study the
critical behavior of BI-AdS black hole while the cosmological constant is
fixed and the electric charge of black hole can vary.
%%%%%%%%%%%%%%%%%%%%%%%%%%%%%%%%%%%%%%%%%%%%%%%%%%%%%%%%%%%%%%%%%%%%%%%%%%%%%%

\section{Critical behavior of BI-AdS black hole \label{PT}}

\begin{figure}[t]
\epsfxsize=8.5cm \centerline{\epsffile{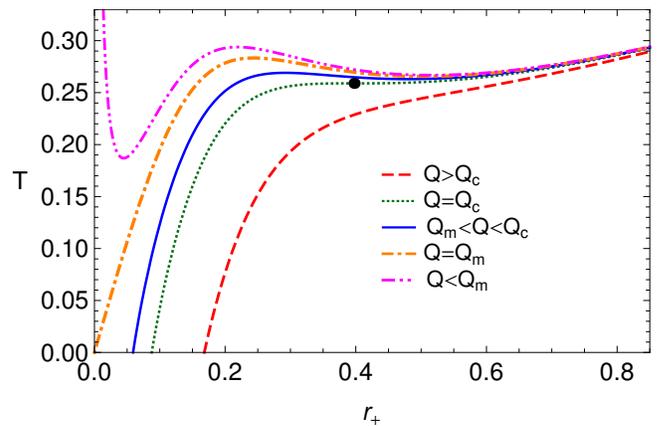}}
\caption{$T-r_{+}$ diagram of BI-AdS black hole. This figure shows the
different behavior depending on the charge of black hole. A negative slope
appears for $Q<Q_{c}$ which indicates the local instability ($C_{Q}<0$). The
critical point is indicated by a black spot. We have set $L=1$ and $\protect%
\beta =3.5$.}
\label{fig1}
\end{figure}

In this section, we investigate the critical behavior of BI-AdS black hole
in which the charge of black hole can vary whereas the cosmological constant
is a fixed parameter. In order to do this, we can study the behavior of the
specific heat at constant charge%
\begin{equation}
C_{Q}=T\left( \frac{dS}{dT}\right) _{Q},  \label{CQ}
\end{equation}%
where, the positive (negative) sign of this quantity indicates the local
thermodynamic stability (instability). Note that Eq. (\ref{CQ}) is also
calculated at fixed $L$ (i.e. cosmological constant) and $\beta $. To see
the behavior of $C_{Q}$, we plot the curves of the temperature versus the
event horizon for different values of the charge in Fig. \ref{fig1}. As one
can see, for small $r_{+}$, the behavior of the temperature strongly depends
on the charge of the black holes. Hence, we expand the Hawking temperature
for small $r_{+}$ as
\begin{equation}
T=\frac{2\beta }{r_{+}}\left( Q_{m}-Q\right) +\frac{r_{+}\left(
3+2L^{2}\beta ^{2}\right) }{4\pi L^{2}}-\frac{\beta ^{3}r_{+}^{3}}{16\pi
^{2}Q}+\mathcal{O}\left( r_{+}^{4}\right) ,
\end{equation}%
where $Q_{m}=1/\left( 8\pi \beta \right) $ is the `\textit{marginal }%
charge'. Also, the large $r_{+}$ limit of the temperature is $3r_{+}/\left(
4\pi L^{2}\right) $ which is independent of the charge. Hence for large $%
r_{+}$, the temperature increases linearly with increasing $r_{+}$.
Depending on the value of $\beta $, BI-AdS black hole may be identified as
follows: For $Q\geq Q_{m}$, black hole is RN type. In this case, we have an
extremal black hole because temperature crosses over from zero with
decreasing $r_{+}$. For $Q<Q_{m}$, black hole is `\textit{Schwarzschild-like}%
' (S) type. Similar to Schwarzschild solution, black hole does not exist in
the region of low temperature. According to Fig.\ref{fig1}, the rightmost
(leftmost) branch of isocharge is related to large black hole which is
locally stable (unstable), i.e. positive (negative) sign of specific heat at
constant charge.

Now, we turn to obtain the critical point (where continuous phase transition
occurs) with respect to the above mentioned for BI-AdS black hole. In Fig. %
\ref{fig1}, isocharge diagrams show that, for constant $L$ and $Q=Q_{c}$,
the critical point is an inflection point which can be characterized by%
\begin{equation}
\frac{\partial T}{\partial r_{+}}\Big|_{Q_{c}}=0,\quad \quad \quad \frac{%
\partial ^{2}T}{\partial r_{+}^{2}}\Big|_{Q_{c}}=0.  \label{cpoint}
\end{equation}%
Note that for S-type black hole, the critical point must be occurred in the
right branch of $Q_{c}$ which is locally stable. Using the temperature
formula in Eq.(\ref{Tem}), the expressions of Eq. (\ref{cpoint}) are,
respectively, written as%
\begin{eqnarray}
\frac{1}{x}-2x+\left( 1+\frac{3}{2\beta ^{2}L^{2}}-\frac{1}{2\beta
^{2}r_{+c}^{2}}\right) &=&0,  \label{cp1} \\
x^{3}-\frac{x}{2}-\frac{1}{2x}+\frac{1}{4\beta ^{2}r_{+c}^{2}} &=&0,
\label{cp2}
\end{eqnarray}%
where
\begin{equation}
x=\left( \sqrt{1+\frac{16\pi ^{2}Q_{c}^{2}}{\beta ^{2}r_{+c}^{4}}}\right)
^{-1}.
\end{equation}%
Since the critical quantities are real positive values, we should have the
following constraint on $x$,
\begin{equation}
0\leq x\leq 1.  \label{constraint}
\end{equation}

Combining Eqs. (\ref{cp1}) and (\ref{cp2}) gives the cubic equation%
\begin{equation}
x^{3}+px+q=0,
\end{equation}%
where%
\begin{equation}
p=-\frac{3}{2},\quad q=\frac{1}{2}\left( 1+\frac{3}{2\beta ^{2}L^{2}}\right)
.
\end{equation}%
Due to fact that $q$ is real and $p<0$, the cubic equation has three or one
real roots. Three roots exist if $4p^{3}+27q^{2}\leq 0$, i.e.%
\begin{equation}
\beta \geq \beta _{0}=\sqrt{\frac{3}{2}\left( 1+\sqrt{2}\right) }/L\approx
1.9098/L.
\end{equation}%
In this case, the roots are written as%
\begin{eqnarray}
&&x_{k}=\sqrt{2}\mathrm{cos}\left( \frac{1}{3}\mathrm{arccos}\left[ -\frac{%
\sqrt{2}}{2}\left( 1+\frac{3}{2\beta ^{2}L^{2}}\right) \right] -\frac{2\pi k%
}{3}\right) ,  \notag \\
&&k=0,1,2.
\end{eqnarray}%
One can easily verify that the constraint in Eq.(\ref{constraint}) is
violated by $x_{2}$. Also, by calculating the critical quantities from $%
x_{1} $, we find that $r_{+c}$ occurs in the unstable branch of critical
isocharge for S-type black hole. For $\beta <\beta _{0}$ case, one root is
given by%
\begin{equation}
x_{3}=-\sqrt{2}\mathrm{cosh}\left( \frac{1}{3}\mathrm{arccosh}\left[ -\frac{%
\sqrt{2}}{2}\left( 1+\frac{3}{2\beta ^{2}L^{2}}\right) \right] \right) ,
\end{equation}%
which violates the condition in Eq. (\ref{constraint}). Consequently, the
critical behavior of BI-AdS black hole can be observed only for $\beta \geq
\beta _{0}$. With $x_{0}$ at hand, the critical quantities are obtained as%
\begin{eqnarray}
&&r_{+c}^{-2}=2\beta ^{2}\left( \frac{1}{x_{0}}+x_{0}-2x_{0}^{3}\right) , \\
&&Q_{c}^{-2}=64\pi ^{2}\beta ^{2}\left( 1-x_{0}^{2}\right) \left(
1+2x_{0}^{2}\right) ^{2}, \\
&&T_{c}=\frac{\sqrt{x_{0}}\left[ 3+2L^{2}\beta ^{2}\left( 1-x_{0}\right)
\left( 1+2x_{0}+2x_{0}^{2}\right) \right] }{4\pi L^{2}\beta \sqrt{2\left(
1-x_{0}^{2}\right) \left( 1+2x_{0}^{2}\right) }}
\end{eqnarray}%
The critical charge is larger than $Q_{m}$ when%
\begin{equation}
\beta >\beta _{1}=\sqrt{\frac{3}{2(\sqrt{6\sqrt{3}-9}-1)}}/L\approx 2.8871/L.
\end{equation}%
Thus for $\beta \geq \beta _{1}$, the critical behavior takes place in the
RN-type of black hole. On the other hand, for $\beta _{0}\leq \beta <\beta
_{1}$, the critical point occurs for S-type black hole at the right branch
of $Q=Q_{c}$.

Expanding the critical quantities of BI-AdS black hole for large $\beta $
(RN-type), lead to
\begin{eqnarray}
r_{+c} &=&\frac{L}{\sqrt{6}}-\frac{7}{24\sqrt{6}\beta ^{2}}+\mathcal{O}%
\left( \frac{1}{\beta ^{4}}\right) ,  \notag \\
Q_{c} &=&\frac{L}{24\pi }+\frac{7}{576\pi L\beta ^{2}}+\mathcal{O}\left(
\frac{1}{\beta ^{4}}\right) ,  \notag \\
T_{c} &=&\frac{1}{L\pi }\sqrt{\frac{2}{3}}-\frac{1}{12\sqrt{6}\pi L^{3}\beta
^{2}}+\mathcal{O}\left( \frac{1}{\beta ^{4}}\right) .  \label{expandcp}
\end{eqnarray}%
As expected, the first terms on the right-hand side of Eqs.(\ref{expandcp})
reproduces the critical values of RN-AdS black hole \cite{AAA}. The leading
BI correction to critical point is identified by the last terms in above
equations. To determine the possible phase transition in the system, we
study the Gibbs free energy in the next section.
%%%%%%%%%%%%%%%%%%%%%%%%%%%%%%%%%%%%%%%%%%%%%%%%%%%%%%%%%%%%%%%%%%%%%%%

\section{Gibbs free energy\label{Gibbs}}

To find out phase transition and classify its type, we need to explore the
behavior of thermodynamic potential (partition function) of BI-AdS black
hole. This is due to the fact that the thermodynamic potential determines
the globally stable state at equilibrium process. For fixed temperature $T$,
pressure $P$\ ($\Lambda $) and charge $Q$, the Gibbs free energy is
themodynamic potential which is computed from the Euclidean action with
appropriate boundary term in the canonical ensemble \cite{PV,BIRPT}. In this
case, such a global stable state is corresponding to the lowest Gibbs free
energy. Since the mass of black hole is identified as the enthalpy Eq. (\ref%
{enthalpy}), one obtains the Gibbs free energy per unit volume $\omega $ by
Legendre transformation \cite{BIRPT}%
\begin{eqnarray}
&&G\left( T,Q,P\right) =M-TS=\frac{r_{+}}{16\pi }-\frac{r_{+}^{3}}{16\pi
L^{2}}-\frac{r_{+}^{3}\beta ^{2}}{24\pi }  \notag \\
&&\times \left( 1-\sqrt{1+\frac{16\pi ^{2}Q^{2}}{\beta ^{2}r_{+}^{4}}}%
\right) +\frac{8\pi Q^{2}}{3r_{+}}\mathcal{F}\left[ \frac{1}{4},\frac{1}{2},%
\frac{5}{4},-\frac{16\pi ^{2}Q^{2}}{\beta ^{2}r_{+}^{4}}\right] ,  \notag \\
&&  \label{GFE}
\end{eqnarray}%
in which $r_{+}=r_{+}(T,Q,L)$\ and $L=L(P)$. In the following, we
shall assume the charge of BI-AdS black hole can vary (parametric
changes in the Gibbs free energy in Eq.(\ref{GFE})), while the
pressure do not take different values. {It is worth mentioning
that since the pressure and }$\beta ${ are fixed, one may consider
the above equation as the Helmhotz free energy. However, the
results of this section are not changed.}

\begin{figure}[t]
\epsfxsize=8.5cm \centerline{\epsffile{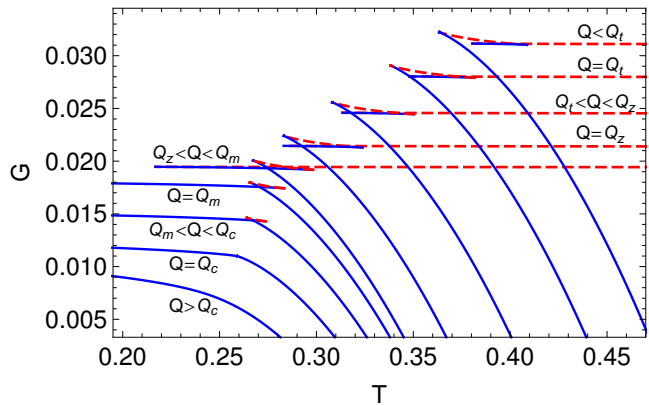}}
\caption{Gibbs free energy as a function of temperature for different values
of charge and $\protect\beta \geq \protect\beta _{1}$. For $Q\geq Q_{m}$,
the phase transition behavior is reminiscent of RN black hole. Reentrant
phase transition (LBH/SBH/LBH) takes place for $Q_{t}\leq Q\leq Q_{z}$. For $%
Q<Q_{t}$, the lower (upper) branch is globally stable (unstable). The
positive (negative) sign of $C_{Q}$ is denoted by the blue solid (dash red)
line. Note that curves are shifted for clarity. We fix $L=1$ and $\protect%
\beta =3.5$.}
\label{fig2}
\end{figure}
\begin{figure}[t]
\epsfxsize=8.5cm \centerline{\epsffile{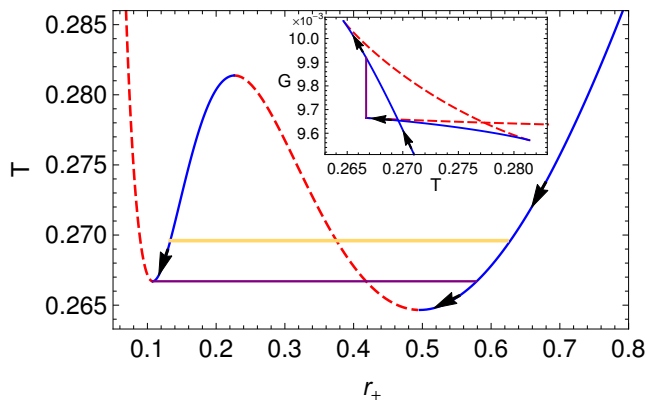}}
\caption{The behavior of isocharge $T-r_{+}$ diagram and the corresponding $%
G-T$ diagram (inset) of BI-AdS for the case of $L=1$, $Q=0.0101$ and $%
\protect\beta =3.5$. As temperature decreases the black hole follows
direction of arrows. The first and zeroth order phase transition are
identified by gold and purple curve, respectively. A
large/small/large(intermediate) corresponds to a reentrant phase transition.
The positive (negative) sign of $C_{Q}$ is denoted by the blue solid (dash
red) line.}
\label{fig3}
\end{figure}
\begin{figure}[t]
\epsfxsize=8.5cm \centerline{\epsffile{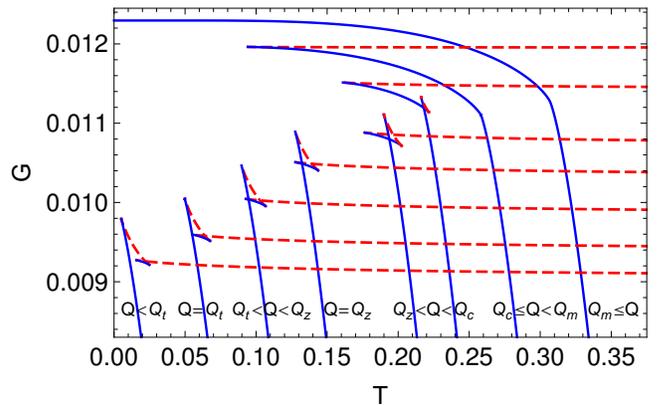}}
\caption{Gibbs free energy as a function of temperature for different values
of charge and $\protect\beta $ $\in \left[ \protect\beta _{0},\protect\beta %
_{1}\right) $. For $Q\geq Q_{m}$, there is no phase transition in the
system. Similar to the schwarzschild-AdS black hole, for $Q_{c}<Q<Q_{m}$,
the lower branch (LBH) is thermodynamically stable, while the upper branch
(SBH) is unstable. The first order phase transition between small and large
black holes only occurs in range of $Q_{z}<Q<Q_{c}$. Same as the Fig.(%
\protect\ref{fig2}) in between $Q_{t}\leq Q\leq Q_{z}$, there is a reentrant
phase transition LBH/SBH/LBH. There is no phase transition below $Q=Q_{t}$.
The positive (negative) sign of $C_{Q}$ is denoted by the blue solid (dash
red) line. Note that curves are shifted for clarity. We fix $L=1$ and $%
\protect\beta =2.83$.}
\label{fig4}
\end{figure}

The behavior of Gibbs free energy for $\beta \approx 3.5$ $\in \left[ \beta
_{1},\infty \right) $ and $\beta \approx 2.84$ $\in \left[ \beta _{0},\beta
_{1}\right) $ are illustrated in Figs. \ref{fig2} and \ref{fig4},
respectively. As one can see from Figs. \ref{fig2} and \ref{fig4}, BI-AdS
black hole has a different and complex behavior for various charge $Q$. In
Fig. \ref{fig2}, there is a critical point at $Q=Q_{c}$ in the region of
RN-type. For $Q>Q_{c}$, the Gibbs free energy is single valued and is
locally stable ($C_{Q}>0$) every where, which is shown by the solid blue
line. However, for $Q<Q_{c}$, it becomes multi valued in which negative $%
C_{Q}$ identified by dash red line. In the range of $Q_{z}<Q<Q_{c}$, a first
order phase transition occurs between small black hole (SBH) and large black
hole (LBH) which is accompanied by a discontinuity in the slop of $G$ at
transition point. Also, for $Q<Q_{m}$, black hole becomes a S-type where
temperature of black hole has a lower bound. An interesting phenomenon
emerges in the range of $Q_{t}<Q<Q_{z}$ where the reentrant phase transition
occurs between intermediate (large) black hole, SBH and LBH in S-type
region. This means that in addition to the first order phase transition
which separates SBH and LBH, a finite jump in Gibbs free energy leads to a
zeroth order phase transition between SBH and intermediate black hole (LBH)
where initiates from $Q=Q_{z}$ and terminates at $Q=Q_{t}$. The zeroth order
phase transition earlier seen in superfluidity and superconductivity in \cite%
{maslov} and recently has been found in charged dilaton black hole \cite%
{NAAA}. Finally, for $Q<Q_{t}$, only LBH exists that is stable.

Especially, let us consider the case of $Q\approx 0.0101$ $\in \left[
Q_{t},Q_{z}\right] $ in Fig. \ref{fig3}. Decreasing the radius of the event
horizon, black hole follows the lower solid blue curve until it joints to
the upper solid blue curve in $G$-$T$ plane (see the inset in Fig. \ref{fig3}%
). In this position, black hole enters to the left solid blue curve with a
first order LBH/SBH phase transition which is identified by gold line in $T$-%
$r_{+}$ plane. If we continue decreasing $r_{+}$, a zeroth order phase
transition occurs between SBH and intermediate black hole (LBH) which is
accompanied by a finite jump in $G$ at the end of solid blue curve. The
zeroth order phase transition is displayed by purple line in Fig. \ref{fig3}%
. Finally, with decreasing $r_{+}$, black hole follows the solid blue line
until the end. In case of $\beta \approx 2.84$ $\in \left[ \beta _{0},\beta
_{1}\right) $, Fig. \ref{fig4}, the behavior of Gibbs free energy is similar
to the mentioned above while all the phase transition takes place in region $%
Q<Q_{m} $, i.e. S-type black hole.
\begin{figure}[t]
\epsfxsize=8.5cm \centerline{\epsffile{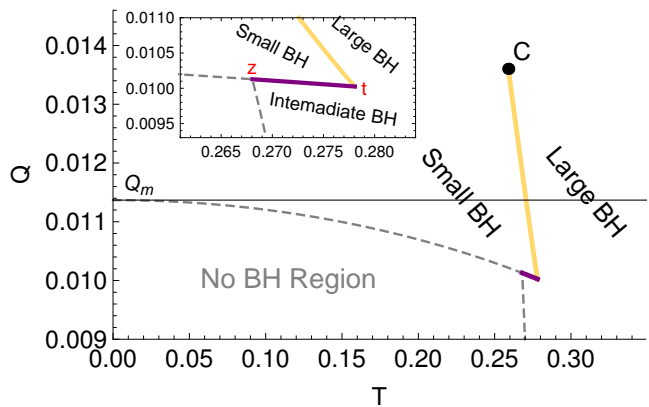}}
\caption{Phase diagram corresponding to Fig.(\protect\ref{fig2}). Three
different transition are indicated: zeroth-order (purple curve), first-order
(gold curve) and a second-order (black solid circle). The critical point
happens in the RN-type black hole. The inset demonstrates the presence of a
large/small/large (intermediate) black hole reentrant phase transition for a
small range of charge. The marginal charge is shown by black horizontal
line. At charges below the marginal charge, no BH region corresponds to no
black hole solution for region of parameter space.}
\label{fig5}
\end{figure}
\begin{figure}[t]
\epsfxsize=8.5cm \centerline{\epsffile{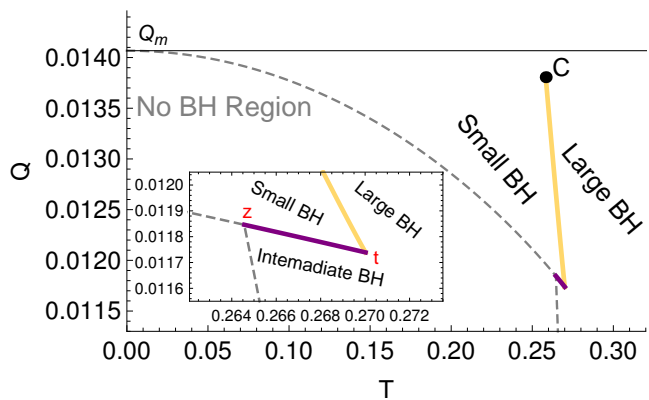}}
\caption{Phase diagram corresponding to Fig.(\protect\ref{fig4}). Three
different transition are indicated: zeroth-order (purple curve), first-order
(gold curve) and a second-order (black solid circle). Phase transition
configuration emerges in the S-type black hole, i.e. below the marginal
charge. The inset demonstrates the presence of a large/small/large
(intermediate) black hole reentrant phase transition for a small range of
charge. The marginal charge is shown by black horizontal line. At charges
below the marginal charge, no BH region corresponds to no black hole
solution for region of parameter space.}
\label{fig6}
\end{figure}
The corresponding phase diagram of BI-AdS black hole for $\beta \approx 3.5$
$\in \left[ \beta _{1},\infty \right) $ and $\beta \approx 2.84$ $\in \left[
\beta _{0},\beta _{1}\right) $ are depicted in the $Q$--$T$ diagram in Figs. %
\ref{fig5} and (\ref{fig6}), respectively. As is seen from Fig. \ref{fig5},
the critical point is highlighted by a black spot and the first order phase
transition separation small and large black hole in the region of $%
Q_{t}<Q<Q_{c}$ which is marked by the gold curve. For $Q_{t}<Q<Q_{z}$, along
with the fist order SBH/LBH, a zeroth order phase transition separates
SBH/LBH (SBH/IBH) which is identified by purple curve. Hence, we have
LBH/SBH/LBH phase transition in the region of $Q_{t}<Q<Q_{z}$, which is the
reentrant phase transition. There is no phase transition in the charge range
$Q>Q_{c}$ as well as $Q<Q_{t}$. Also, no BH region corresponds to no black
hole solution exist. Fig.(\ref{fig6}) shows different phase of BI-AdS black
hole for $\beta \approx 2.84$ $\in \left[ \beta _{0},\beta _{1}\right) $
which is similar to Fig. \ref{fig5} while zeroth, first and second order
phase transition occur in S-type of black hole. An important point can be
understood from above discussion: the reentrant phase transition emerges for
the S-type of black hole, i.e. the charge of black hole is less than
marginal charge.
%%%%%%%%%%%%%%%%%%%%%%%%%%%%%%%%%%%%%%%%%%%%%%%%%%%%%%%%%%%%%%%%%%%%

\section{Summary\label{conclusion}}

To summarize, we have investigated charged black holes in the background of
AdS spacetime and in the presence of BI nonlinear electrodynamics. We have
proposed a new viewpoint towards phase space of black holes by assuming that
the cosmological constant, which is regard as the pressure of the system, is
a fixed quantity, while in contrast the charge of the black hole can vary.
Recently, it was argued that such an alternative view towards phase space of
charged AdS black holes can successfully lead to a natural response function
which measures how the size of a black hole ($\Psi \sim r_{+}^{-1}$) changes
with its charge. It was shown that this alternative view more importantly
leads to a natural correspondence of the charged AdS black hole with the Van
der Waals fluid and the associated small-large black hole phase transition
without a need for extended phase space \cite{AAA}.

First, we explored critical behavior of BI-AdS black holes by studying the
behavior of the specific heat at constant charge, $C_{Q}$, which its sign
indicates the local thermodynamic stability/instability of the system. We
find that for small horizon radius $r_{+}$, the behavior of the temperature
strongly depends on the charge of the black holes. We observed that
depending on the value of the nonlinear parameter, $\beta $, BI-AdS black
hole may be identified as RN black hole for $Q\geq Q_{m}$, and
Schwarzschild-like black hole for $Q<Q_{m}$, where $Q_{m}=1/\left( 8\pi
\beta \right) $ is the \textit{marginal} charge. Besides, similar to
Schwarzschild solution, black hole does not exist in the region of low
temperature. We analytically calculated the critical point ($Q_c,T_c, r_{+c}$%
) by solving the cubic equation and studied the critical behavior of the
system.

We have also explored the behavior of Gibbs free energy for BI-AdS black
hole. We find out that BI-AdS black hole has a different and complex phase
behavior depending on the value of the charge $Q$ of the system. Indeed,
there is a critical point at $Q=Q_{c}$ in the region of RN-type black holes.
For $Q>Q_{c}$, the Gibbs free energy is single valued and the system is
locally stable ($C_{Q}>0$). However, for $Q<Q_{c}$, we have $C_{Q}<0$ and
the system experiences thermally instable phase. We observed that for $%
Q_{z}<Q<Q_{c}$, a first order phase transition occurs between SBH and LBH
which is accompanied by a discontinuity in the slop of Gibbs free energy at
transition point. An interesting phenomenon emerges for $Q$ in the range of $%
Q_{t}<Q<Q_{z}$, where a reentrant phase transition can be occurred between
intermediate (large) BH, SBH and LBH in Schwarzschild-type region. This
implies that in addition to the first order phase transition which separates
SBH and LBH, a finite jump in Gibbs free energy leads to a \textit{zeroth
order} phase transition between SBH and intermediate black hole (LBH) where
initiates from $Q=Q_{z}$ and terminates at $Q=Q_{t}$.

Finally, it is important to mention that in the present work we only studied
the phase behavior of BI-AdS black holes. This investigation can also be
extended to other BI-like nonlinear electrodynamics such as Exponential and
Logarithmic nonlinear electrodynamics. Besides, it is worth disclosing the
effects of power parameter of the power-Maxwell electrodynamics on the
reentrant phase transition of black holes in the background of AdS spaces
with variable charge. These issues are now under investigation and the
results will be appeared in the near future.
%%%%%%%%%%%%%%%%%%%%%%%%%%%%%%%%%%%%%%%%%%%%%%%%%%%%%%%%%%%%%%%%%%%%%%%%%%%%%%

\begin{acknowledgments}
We acknowledge the referees for constructive comments which helped us
improve our paper significantly. We also grateful the Research Council of
Shiraz University. This work has been supported financially by Research
Institute for Astronomy \& Astrophysics of Maragha (RIAAM), Iran.
\end{acknowledgments}

%%%%%%%%%%%%%%%%%%%%%%%%%%%%%%%%%%%%%%%%%%%%%%%%%%%%%%%%%%%%%%%%%%%%%%%%%%%%%%%%

\end{document}